# Creating Cybersecurity Regulatory Mechanisms, as Seen Through EU and US Law

*Kaspar Rosager Ludvigsen* (Durham University, UK)

**Abstract**

Because digital devices and systems are widely used in all aspects of society, the risk of adversaries creating cyberattacks on a similar level remains high. As such, regulation of these aspects must follow – which is the domain of cybersecurity. Because this topic is worldwide, different jurisdictions should take inspiration from successful techniques elsewhere, with the European Union (EU) and the US being the most experienced and long-standing. What can be derived from their approaches separately to be used in other democratic jurisdictions, and what happens when we compare them with this pragmatic approach in mind? Cybersecurity is oddly enough quite well understood in most jurisdictions worldwide. However, concept comprehension cannot enforce or create compliance, hence the need for good regulatory approaches. The comparative legal analysis of the EU and the US show that there are large differences in definitions and enforcement, but some concepts are repeated in both jurisdictions. These can be further refined to become derivable principles, which can be used to inspire legislation in any democratic jurisdiction. They are: Voluntary Cooperation, Adaptable Definitions, Strong-arm Authorities, Mandated Computer Emergency Response Teams, and Effective Sanctions. These 5 principles are not exhaustive but combine classic regulatory and practical lessons from these two jurisdictions.

## I. Introduction

Cybersecurity is the essential defence in digital systems, and is mandatory to prevent other people, government entities, machines, and others (collectively known as adversaries) to access, control, or destroy digital systems or devices. Cybersecurity can be negatively defined as a state of peace and lack of cyberattacks succeeding (Zdzikot 2022, 17–18). Some jurisdictions have extensive positively written and mandated legislation, which require cybersecurity from private and state actors, or which lets states supervise or support companies or individuals in attaining adequate cybersecurity, such as Poland (Brzostek 2022), Germany (Martini and Kemper 2022; Schallbruch and Skierka 2018) or the US (Lessambo 2023; Lubin 2023; Fischer 2013). However, this is not the case everywhere, and those jurisdictions who lack the understanding or implementation of the regulation of cybersecurity is whom this paper is written for. Considering what existing legal systems have done, through comparative legal analysis, is the tool to seek inspiration for new legislation in any given country.

We therefore set out to give inspiration to the legal systems who have yet to take cybersecurity into their own hands and mandate the necessary technical requirements and powers to the right authorities. We do so through comparative legal analysis (this is also the methodology) of two large legal systems, the European Union (EU) and the US. These are large and influential legal systems globally and have regulated cybersecurity for many years. From this analysis, we derive 5 principles which could be used to design custom cybersecurity regulation in jurisdictions which lack them. These are: Voluntary Cooperation, Adaptable Definitions, Strong-arm Authorities, Mandated Computer Emergency Response Teams, and Effective Sanctions.

**II. Cybersecurity Regulation in EU and US Law**

Before making any recommendations, comparative legal analysis is necessary to show where other jurisdictions are regarding the regulation of cybersecurity. This consists of characterising the systems and include references to the relevant legislation and policy,[1] which for both US and EU law is freely available online on websites hosted by both jurisdictions. We start with the EU.

*EU*

EU is not just one legal system; it sits above the jurisdictions of each of the 27 Member States (MS) and is therefore what we could call a "supranational" (Canihac 2020; Gabriel 2019) legal system.[2]

The EU regulates cybersecurity in its product legislation, its separate specialised rules, critical infrastructure legislation, and indirectly in legal acts such as the GDPR.[3] The EU cannot legislate directly on areas such as cybercrime, or criminal elements of cybersecurity otherwise, as this is limited by the Treaty of the Functioning of the European Union, where the EU does not have sole or shared competence on the topic of criminal law,[4] which is required for it to legislate and regulate the area.

---

[1] Further in-depth comparative legal analysis is outside the scope of this paper, additional research by the author (Ludvigsen 2023; Ludvigsen and Nagaraja 2022b; 2022a) or others (van 't Schip 2024; Chiara 2023; Carr and Tanczer 2018; Casarosa 2022) should be consulted for this instead.

[2] This refers to legal systems above others, which are not international or global, such as the African Union.

[3] Regulation (EU) 2016/679 of the European Parliament and of the Council of 27 April 2016 on the protection of natural persons with regard to the processing of personal data and on the free movement of such data, and repealing Directive 95/46/EC (General Data Protection Regulation) [2016] OJ L 1191

[4] See Art 3 and 4 of the Treaty of the Functioning of the European Union.

Specialised rules include the Cybersecurity Act (CA),[5] and the Cyber Solidarity Act (CSA),[6] while product legislation covers everything from e.g., those regulating Medical Devices, the Medical Device Regulation (MDR),[7] to the AI Act[8] and the Cyber Resilience Act (CRA).[9]

EU has cooperative mechanisms between big private cybersecurity providers (Bossong and Wagner 2017), or big structural companies such as Meta, which exist in practice, but will only be explicitly mentioned in the upcoming CSA. It has a central authority in the form of the European Union Agency for Cybersecurity (ENISA), empowered in the CA, though MS authorities play the largest role, meaning these are all relevant and important both currently and going forward. Critical Infrastructure includes the defences of all systems which are necessary for the functioning of any MS, meaning that NIS1[10] and NIS2[11] must be included. Sadly, the NIS1 implementation was considered to be fragmented and lacklustre (Ludvigsen, Nagaraja, and Daly 2022; Wallis and Johnson 2020; Kelemen, Szabo, and Vajdová 2018).

Cybersecurity in product legislation drives practical implementation across sectors and by private actors, who then must expect to be inspected or tested by relevant authorities, where security is usually regulated through guidance or indirect wording or interpretation of the primary product legislation like the MDR (Ludvigsen and Nagaraja 2022a; Biasin and Kamenjasevic 2020).[12] Or through all types of systems with network connectivity, which applies through the proposed CRA, meaning all products that provide such functionality much fulfil its minimum cybersecurity requirements across the EU (Eckhardt and Kotovskaia 2023).

---

[5] Regulation (EU) 2019/881 of the European Parliament and of the Council of 17 April 2019 on ENISA (the European Union Agency for Cybersecurity) and on information and communications technology cybersecurity certification and repealing Regulation (EU) No 526/2013 (Cybersecurity Act), OJ L151/15.
[6] Proposal for a Regulation of the European Parliament and of the Council laying down measures to strengthen solidarity and capacities in the Union to detect, prepare for and respond to cybersecurity threats and incidents, COM(2023) 209 final, 2023/0109 (COD).
[7] Regulation (EU) 2017/745 of the European Parliament and of the Council of 5 April 2017 on medical devices, amending Directive 2001/83/EC, Regulation (EC) No 178/2002 and Regulation (EC) No 1223/2009 and repealing Council Directives 90/385/EEC and 93/42/EEC [2017] OJ L117/1.
[8] REGULATION (EU) OF THE EUROPEAN PARLIAMENT AND OF THE COUNCIL, laying down harmonised rules on artificial intelligence and amending Regulations (EC) No 300/2008, (EU) No 167/2013, (EU) No 168/2013, (EU) 2018/858, (EU) 2018/1139 and (EU) 2019/2144 and Directives 2014/90/EU, (EU) 2016/797 (Final Draft).
[9] Proposal for a Regulation of the European Parliament and of the Council on horizontal cybersecurity requirements for products with digital elements and amending Regulation (EU) 2019/1020, COM(2022) 454 final, 2022/0272 (COD).
[10] Directive (EU) 2016/745 of the European Parliament and of the Council of 6 July 2016 concerning measures for a high common level of security of network and information systems across the Union, OJ L194/1.
[11] Directive (EU) 2022/2555 of the European Parliament and of the Council of 14 December 2022 on measures for a high common level of cybersecurity across the Union, amending Regulation (EU) No 910/2014 and Directive (EU) 2018/1972, and repealing Directive (EU) 2016/1148 (NIS 2 Directive), [2022] OJ L333/80.
[12] This should continue through mechanisms such as post-market surveillance (Badnjević et al. 2022; Zippel and Bohnet-Joschko 2017), where medical devices are a good example as well, even in a cybersecurity context.

Cybersecurity matters in product liability situations too (Ludvigsen and Nagaraja 2022a), meaning the old and proposed new Product Liability Directive[13] will be used in situations where litigation is necessary, including security and safety (Buiten 2023, 22–23; Alheit 2001), as well as the proposed AI Liability Directive.[14]

*US*

US law is a wide concept, spanning both Federal law and State law (Hart 1954). We focus on the Federal rules, as making an overview of what each US state has implemented of legislation on top of the number of federal rules is beyond the scope of this paper.[15]

At this level, it exists past legislation which can be considered as cybersecurity regulation, such as the National Security Act of 1947,[16] and product legislation partially from the Consumer Product Safety Act.[17] Unlike EU law, US law excels in its specifications for the cybersecurity for the state as such, which can be seen with e.g., the Electronic Communications Privacy Act of 1986,[18] Economic Espionage Act,[19] Cybersecurity Act 2015[20] (CA2015), and recently, the policy of the National Cyber Strategy,[21] the Cyber Incident Reporting for Critical Infrastructure Act[22] and the Strengthening American Cybersecurity Act 2022.[23] The latter might not solve the fragmented state which some authors see the current cybersecurity landscape in the US being in (Kosseff 2018; 2023), and there is no equivalent all-encompassing general cybersecurity law in the US either yet (Lubin 2023, 23). The potential in having strong connections between different authorities is great and considerable as well (Roesener, Bottolfson, and Fernandez 2014), creating synergies which are worth considering for other legal systems.

Products are further controlled by large central authorities in the form of the Federal Trade Commission (FTC) (Dibert 2016) and others, and there is privacy legislation like the Health Insurance Portability and Accountability Act (Jalali and Kaiser 2018) which involves some cybersecurity criteria. The role of guidance and choices in inspection will decide the cybersecurity level, but this can be a disadvantage as well (Roth 2014), something which is hopefully being changed over time in a security context.

---

[13] Directive of the European Parliament and of the Council on liability for defective products, Brussels, 28.9.2022 COM(2022), 495 final, 2022/0302 (COD).
[14] Directive of the European Parliament and of the Council on adapting non-contractual civil liability rules to artificial intelligence
(AI Liability Directive)28.9.2022 COM(2022), 496 final 2022/0303 (COD).
[15] The same can be said about the amount of specific legislation which exists in US law that regulates cybersecurity, for a good overview, see (Fischer 2013, 4).
[16] Title 50, United States Code (U.S.C), Chapter 15, which is a barrier to sharing cybersecurity information if classified (Fischer 2013, 32). The U.S.C location denotes implementation into the Code.
[17] U.S.C., Title 15, Chapter 47.
[18] U.S.C., Title 18, Chapter 119, 121, 206.
[19] U.S.C., Title 18, Chapter 90.
[20] U.S.C., Title 6, Chapter 6.
[21] https://www.whitehouse.gov/briefing-room/statements-releases/2023/03/02/fact-sheet-biden-harris-administration-announces-national-cybersecurity-strategy/, last accessed 19 July 2024.
[22] U.S.C., Title 6, Chapter 1.
[23] Part of, the Consolidated Appropriations Act, 2022, see https://www.congress.gov/bill/117th-congress/house-bill/2471, last accessed 19 July 2024.

Cooperation with private entities is natural and built into the CA15 (Kosseff 2018, 39), and exists informally as it does in the EU, and the power of other authorities, such as the Department of Homeland Security [24] (Ohm and Kim 2022) in relation to non-product cybersecurity regulation remains high as well. The nature of cybersecurity breaches (adversarial failures), however, in the form of private entities not wanting to share or release information until absolutely forced to, has consequences that warrant further scrutiny (Schwarcz, Wolff, and Woods 2022), and if this is continues to be an issue in US law, must be taken into consideration.

Because of the much more detailed legislation on criminal or unwanted cybersecurity actions, US law serves as a great example of state regulation of permissible security behaviour, such as in U.S.C., Title 6, Chapter 6. Combined with strong authorities, whose guidance and practice are clearly available,[25] makes for a somewhat different yet familiar approach to that of the EU.

**III. Regulatory Mechanisms in Cybersecurity**

There are many ways to potentially regulate an area such as cybersecurity. The following 5 recommendations are taken directly from both systems,[26] and should be used to consider how many jurisdictions, which does not yet have detailed legislation for cybersecurity, could design and make their own.[27] Initially, it must be said that extreme harshness or centralised control can impact innovation, willingness to corporate, or even make companies or individuals who create cybersecurity leave the country (Aggarwal and Reddie 2018; Huang and Li 2018; Lewis 2009). Great care must be taken to not cause worse side-effects from regulating cybersecurity, than leaving it unregulated.

*Voluntary Cooperation*

The relative power big companies hold within cybersecurity is very high (Farrand and Carrapico 2018; Carrapico and Farrand 2021; Moore 2010), as is the importance of communicating with providers of critical infrastructure and so on (Dykstra et al. 2022), even if partially state owned. Therefore, establishing both formal and informal communication paths are central to any kind of information sharing and control of the cybersecurity sphere, which is key to any kind of digital development.

---

[24] E.g., https://www.dhs.gov/sites/default/files/publications/Strategic_Principles_for_Securing, last accessed 19 July 2024.
[25] For example, the FDA publishes cybersecurity guidance for medical devices on its website, such as https://www.fda.gov/regulatory-information/search-fda-guidance-documents/changes-existing-medical-software-policies-resulting-section-3060-21st-century-cures-act, last accessed 19 July 2024.
[26] Each principle has references to where they are inspired from, but it is not always literally.
[27] Such a system could be country like Cambodia, which this issue centres about. But the thoughts of this paper is applicable across the world.

An alternative option is to mandate certain threat indicators and defensive measures be shared to the government, which is required in US law.[28] This is very usable, since big corporations who have an interest in either being supported or assisted to update and patch vulnerable systems, can do so in a synchronous manner with the state. However, it can only be done well if there is little monetary or reputational risk to the private company providing this information, else it the information sharing may be illusory (Schwarcz, Wolff, and Woods 2022).

Inspiration can be taken from the CSA in EU law,[29] and especially from the CA15 in US law, and the focus should be on creating power within ministerial or state organs, who can facilitate the contact and maintain it, and integrate these communication channels into all types of state infrastructure – as cyberattacks can happen anywhere, from government computers, hospital equipment, or school iPads.

***Adaptable Definitions***

Technology changes over time, especially when it comes to software and hardware, meaning that the regulation of cybersecurity should be able to adapt in turn. This can be done in variety of ways, which could be so make the legislation very technology neutral (Reed 2007), which is seen in the proposed CRA from the EU, or in the amount of detailed technology specific legislation (Ohm 2010) there exists in the US related to cybersecurity .[30] A good middle ground could be to have legislation which is continuously improved, with sunset clauses which stipulate that the technology regulated is reviewed within a set time period. This could be beneficial when new techniques such as homomorphic or quantum encryption (Kop 2021) become widespread, or the use of Large Language Models increases in the cybersecurity area, or how we saw in practice that cybersecurity of cloud storage became as important as client-side server storage.

An example from US law, is its approach to how post-quantum cryptography will affect national security, and how systems can be migrated to post-quantum defendable systems:

> *… shall issue guidance on the migration of information technology to post-quantum cryptography, which shall include at a minimum-*

---

[28] U.S.C, Title 6, Chapter 6, §1504.
[29] Inspiration can also be taken from individual Member States of the EU, but this is outside the scope of this paper, and varies significantly.
[30] For an overview of the US situation, see Fischer (2013), though this lacks newer legislation.

> *(A) a requirement for each agency to establish and maintain a current inventory of information technology in use by the agency that is vulnerable to decryption by quantum computers, prioritized using the criteria described in subparagraph (B);*
>
> *(B) criteria to allow agencies to prioritize their inventory efforts; and*
>
> *(C) a description of the information required to be reported pursuant to subsection (b)."*[31]

In essence, how the cybersecurity regulation decides to define the technology and subjects it wants to cover should be written and practiced in a flexible manner, which may include adaptable definitions, or implicitly so, through technology-neutral terms.

***Strong-arm Authorities***

Within the cybersecurity community, there has long been a wish for authorities who both understand the nature of security (Abelson et al. 2015; Anderson and Moore 2006), but also provide sanctions for those who chose to disobey the rules. The latter can cause disruptions within various types of product manufacturers, as development costs are naturally lower if poor cybersecurity, or none, is chosen. Comprehending cybersecurity at a fundamental level is necessary because cybersecurity surrounds us and can cause accidents or failures which can harm individuals (Ludvigsen and Nagaraja 2022a), businesses (Alharbi et al. 2021), and entire nations,[32] or at a complete global scale.[33]

With how ENISA is empowered in the EU,[34] and how the FTC is supposed to be,[35] should serve as inspiration as to how to design such a mechanism in a home jurisdiction. An issue arises with those two, in the form of a lack of legal requirements for the professions and skills of staff.

A solution for this is found in the AI Act, which in Article 70(3), where requirements for staff are positively listed, states:

> *"Member States shall ensure that their national competent authorities are provided with adequate technical, financial and human resources, and with infrastructure to fulfil their tasks effectively under this Regulation. In particular, the national competent authorities shall have a sufficient number of personnel permanently available whose competences and expertise shall include an in-depth understanding of AI technologies, data and data*

---

[31] U.S.C, Title 6, Chapter 6, §1526, part a, 1.
[32] For an Estonian view on this problem, see Ebers and Tupay (2023).
[33] See the Crowdstrike Global Outage as an example, https://www.bbc.co.uk/news/articles/cp4wnrxqlewo, last accessed 19 July 2024.
[34] Primarily use CA as inspiration, as ENISA only has sporadic mentions elsewhere. See the CA, Art 6, 7, and 8, for information about the general powers of the ENISA.
[35] Criticism of the lack of enforcement by the FTC justifies focusing on what it should do, over what it does do (Schwarcz, Wolff, and Woods 2022, 45), should be considered.

*computing, personal data protection, cybersecurity, fundamental rights, health and safety risks and knowledge of existing standards and legal requirements. Member States shall assess and, if necessary, update competence and resource requirements referred to in this paragraph on an annual basis."*

Finally, tightly controlling and monitoring cybersecurity manufacturers and distributors too much can lead to a loss of trust, and thereby monetary or reputational losses for the state in question, and this aspect must be considered as well when designing the regulation.

***Mandated Computer Emergency Response Teams***

Writing about or announcing regulation of cybersecurity is not the same as committing to it in practice. To prevent attacks, create defences, and most important, maintain of all this, teams and large amounts of individuals must be at the disposal of the state in some practical manner. This is core part of both EU (CSA proposal) and US law,[36] and requires resources, available space for them in a branch of the government, and collaboration with companies and other actors who share critical information when an incident occurs.[37]

From the CSA, its purpose is worth showcasing, as it could be very useful for building national cybersecurity legislation for these types of response teams:

*''This Regulation lays down measures to strengthen capacities in the Union to detect, prepare for and respond to cybersecurity threats and incidents, in particular through the following actions:*

*(a) the deployment of a pan-European infrastructure of Security Operations Centres ('European Cyber Shield') to build and enhance common detection and situational awareness capabilities;*

*(b) the creation of a Cybersecurity Emergency Mechanism to support Member States in preparing for, responding to, and immediate recovery from significant and large-scale cybersecurity incidents;*

*(c) the establishment of a European Cybersecurity Incident Review Mechanism to review and assess significant or large-scale incidents.''*[38]

These teams, or whatever structure is chosen, should have their own legal provisions, to ensure their availability and adequate skillsets, and exist both at a state and at a local level, akin to how Security Operation Centres[39] exist at an overarching level in the CSA, and National Security Operation Centres.[40]

---

[36] As seen in the National Preparedness System (which includes cybersecurity related scenarios), U.S.C, Title 6, Chapter 2, §742. Worth noting that such emergency response teams also exist under U.S.C, Title 6, Chapter 6, §1522 and §1523, and but are mostly detailed in policy documents and the like, unlike what is detailed in EU's CSA.
[37] Art 59(4) of the AI Act is also good as inspiration here.
[38] CSA, Art 1(1).

*Effective Sanctions*

Preventing continuous breaches of cybersecurity, either by manufacturers, individuals or groups, or foreign states, requires the right answer (Rusinova and Martynova 2023). First, authorities who can inspect, withdraw, or ban products from a market are necessary.[41] Examples of these kinds of powers can be seen in e.g., the FTC or the FDA in US law, or authorities in the AI Act, CRA, CSA and so forth. Secondly, criminal provisions as well as the needed oversight must be considered as part of the regulation of cybersecurity.[42] As foreign states are strong adversaries, special measures, perhaps in relation to the military, should be considered, and represents international law more than domestic law, as it must be considered alongside the Laws of War. This is because cyberattacks can constitute Acts of War (Gervais 2011), and defending against such actions must be understood in this context, and prepared in the relevant legal system. While it can at times be difficult to decide on how, and why in an international legal context (Pomson 2023), it must be either be built into either legislation and/or policy documents.

## IV. Conclusion

In this paper, we have made a quick overview of how US and EU law regulates cybersecurity and formulated 5 principles inspired by elements from both legal systems. The principles have relevance for any state who has yet to make its own cybersecurity regulation and consider existing and functioning solutions, which can help guide countries like Cambodia or others who lack such laws.[43] They are not exhaustive but give an overview of the areas which must be considered. This could be product regulation, state cybersecurity, criminal and other types of liability, and contingency efforts (in case of war or serious adversarial attacks).

While these can inspire some aspects of the process of rulemaking, the rest should be filled in with the relevant legal culture and needs of the country. For further references, the academic papers and legal sources cited should be read, and relevant national stakeholders (private and state) should be included in the drafting processes.

---

[39] CSA, Art 3.
[40] CSA, Art 4.
[41] Fines tend to not be enough, see the commentary by O'Malley (2009).
[42] Strong inspiration for both misuse, fraud, and specialised cybersecurity criminal law problems, can be found in U.S.C, Title 18, §1030.
[43] Some jurisdictions who may not have literal statutory law that concerns cybersecurity, may use existing security, telecommunication, or similar legislation as a replacement until specialised legal sources are created.